\documentclass{ws-p8-50x6-00} 

\usepackage{url,epsfig}

\begin{document}

\title{Parrondo Strategies for Artificial Traders}

\author{Magnus Boman}

\address{Swedish Institute of Computer Science, Box 1263, 
SE-164 29 Kista, Sweden\\ 
E-mail: mab@sics.se}

\author{Stefan J. Johansson}

\address{Department of Software Engineering and Computer Science,\\ Blekinge Institute of Technology, Box 520, SE-372 25, Ronneby, Sweden\\ 
E-mail: sja@bth.se}

\author{David Lyb\"ack}

\address{Financial Market Systems, OM AB,
SE-105 78 Stockholm, Sweden\\
E-mail: david.lyback@omgroup.com}  

\maketitle
\abstracts{On markets with receding prices, artificial noise traders may consider alternatives to buy-and-hold. By simulating variations of the Parrondo strategy, using real data from the Swedish stock market, we produce first indications of a buy-low-sell-random Parrondo variation outperforming buy-and-hold. Subject to our assumptions, buy-low-sell-random also outperforms the traditional value and trend investor strategies. We measure the success of the Parrondo variations not only through their performance compared to other kinds of strategies, but also relative to varying levels of perfect information, received through messages within a multi-agent system of artificial traders.}


\section{Introduction}
Stock markets to an ever-increasing extent allow for trading controlled by artificial agents, or more generally, program trading. For instance, the Swedish Securities Dealers Association finds that it has no objections to program trading, and already in 1992 declared that only the means to exploiting unlawful quotes manipulation, resulting from program trading, should be controlled\cite{ssda92}. Nasdaq, in a communication to their members write\cite{nasd99}: 
\begin{quote}
Recent events show that the way some stocks are traded is changing dramatically, 
and the change in trading methods may affect price volatility and cause increased 
trading volume. This price volatility and increased volume present new hazards to 
investors, regardless of whether trading occurs on-line or otherwise. 
\end{quote}
In general, stock markets do not apply restrictive policies to program trading.
A primary objective of the market place operator is to promote a high liquidity in the traded instruments.
This can be done through reducing the transaction costs: one typical implicit cost is lack of orders, leading to
wide spreads, or non-existing quotes. The operators thus have reasons to encourage inbound orders.
As long as these are authenticated, and the network can keep up disseminating the market info in a proper fashion
so that the situation stays in line with the overall aim of up-keeping a fair
and orderly market, the operator should have nothing against a large number of valid orders per second being placed by
artificial agents.

Hence, we feel motivated to relate certain theoretical results from physics
to artificial traders of the future. We do not assume markets populated solely by artificial traders, however. 
If we did, we could
move on to claim that the Efficient Market Hypothesis and rational choice theory yield efficient equilibria\cite{luau00}, 
since the vast empirical evidence against such assumptions are directed almost exclusively towards human traders\cite{lu95}. We instead believe that
artificial traders have gradually and almost unnoticeably slipped onto the same markets as human traders, and
we will treat them as speculating noise traders (traders with non-rational expectations and potentially zero intelligence)\cite{deshsuwa91}. Artificial stock markets possibly exhibit volatility (i.e., standard deviation) of a different kind than ordinary excess volatility markets\cite{arholepata97}, as argued, e.g., in the ban of crawlers from the Internet auction site eBay\cite{wo00}. In practice, Internet marketplaces supply information on their acceptance of artificial traders and other softbots in a file named \url{robots.txt}, and on Internet markets that do allow for softbots, their behavior is usually monitored in some way, in order to mitigate the effects of speculation through unconventional methods such as denial-of-service attacks. Program trading has also in general reached a level where flocking behavior worry policy makers\cite{egzi00}. On an artificial stock market, in contrast to an ordinary market\cite{mazh98}, active portfolio management should also incorporate the price dynamics, because of the intense trading. This factor has also led to transaction fee policies being radical on some artificial trader markets. Since significant transaction fees can render the Parrondo strategies described in sections~\ref{Parrstrat} and~\ref{Parrinmarket} below useless, the existence of markets with low or no transaction fees is important to our object. We will consider portfolios on markets with receding prices. We will represent artificial traders as agents in a multi-agent system (MAS), in which agents affect each other's behavior through trusted message passing, as explained in section~\ref{Parrinmarket}. In the MAS setting, variations of Parrondo strategies are then subject to experiments on a simulation testbed, the results of which are reported in section~\ref{simulation}. In the last section, we present directions for future research.

\section{The Parrondo Strategy in Games}
\label{Parrstrat}
The flashing ratchet (or Brownian motor)\cite{ajpr92} is a molecular motor system consisting of Brownian particles moving in asymmetric potentials, subject to a source of non-equilibrium\cite{pa98}. 
In its game-theoretical formulation\cite{haab99}, the flashing ratchet can be described in terms of two games (A and B) in which biased coins are tossed. 
\begin{itemize}
\item
Game A is a single coin game in which the coin comes up heads (=win) $50-\epsilon$ per cent of the time (for some small $\epsilon>0$) and results in tails the rest of the times (Parrondo himself\cite{pa98} used $\epsilon = 0.005$, and the constraints are described, e.g., at \url{seneca.fis.ucm.es/parr/GAMES/discussion.html}).
\item
Game B involves two coins. The first coin comes up heads $10-\epsilon$ per cent of the time, and the second coin $75-\epsilon$ per cent of the time. What coin to flip is decided through looking at the capital of the player. If it is divisible by 3, the first coin is flipped, while the second coin is used in the rest of the cases. 
\end{itemize}
Clearly, game A is a losing game, but the same holds for game B. 
This is because the player is only allowed to flip the second coin if her capital is not a multiple of 3. The latter situation comes up more often than every third time: The player will start with the unfavorable coin, which will very likely place her in loss -1. She will then typically revert to 0, and then back again to -1, and so on. Whenever the unfavorable coin lands tails twice in succession, however, she will end up with capital -3, and then the pattern will repeat, leading to -6, etc. Hence, game B is a losing game, just like game A. 

The Parrondo strategy for playing games A and B repeatedly is to choose randomly which game to play next. Somewhat counter-intuitively, this discrete representation of a ratchet yields a winning game.

\section{The Parrondo Strategy in Artificial Trading}
\label{Parrinmarket}
Artificial trading and herd behavior have often been studied through bottom-up simulations, as in Sugarscape\cite{epax96} or the Santa Fe artificial stock market\cite{arholepata97}. We have concentrated on speculating investors that use variations of the Parrondo strategy. Table~\ref{strategytable} briefly describes these strategies, as well as some control strategies. Value investors (exemplified by BLSH in Table~\ref{strategytable}) seek profits, while trend investors (exemplified by BHSL in Table~\ref{strategytable}) try to identify upward and downward movers and adjust their portfolios accordingly\cite{jehahujo00}. In our simulations, the value investor proportion is larger, but this significant fact notwithstanding, our object is not the study of how it affects the market dynamics. Instead, we augment the Parrondo variations by market information, in the form of agent messages. The agents may thus influence each other by passing hints on what to buy, or what to sell. A message is treated by the receiver as trusted information, and the receiving agent will act upon the content of the message, interpreting it as normative advice. A message can be interpreted as perfect (or even insider) information, randomized for the sake of our experiment. 


\begin{table}
\begin{tabular}{|p{0.25\linewidth}|p{0.68\linewidth}|}
\hline
Strategy & Description\\
\hline
\hline
Buy-and-hold (BaH) & The buy-and-hold strategy here acts as a control strategy that trades no stocks.\\
\hline
Random & This strategy trades stocks randomly.\\
\hline
Insider & The insider gets quality ex ante information about some stocks on which it may react before the market.\\
\hline
Buy low, sell high (BLSH) & This Markovian value investor strategy monitors if the stock increased or decreased in value during the latest time interval. If the value increased, it sells the stock, and if the value dropped, it buys the stock.\\
\hline
Buy low, sell random (BLSR) & Like BLSH, except BLSR randomly chooses what stock to sell.\\
\hline
Buy random, sell high (BRSH) & Like BLSH, except BRSH randomly chooses what stock to buy.\\
\hline
Buy high, sell low (BHSL) & This Markovian trend investor strategy is the opposite of BLSH.\\
\hline
\end{tabular}
\caption{The artificial trading strategies.}
\label{strategytable}
\end{table}

\begin{figure}
\epsfig{file=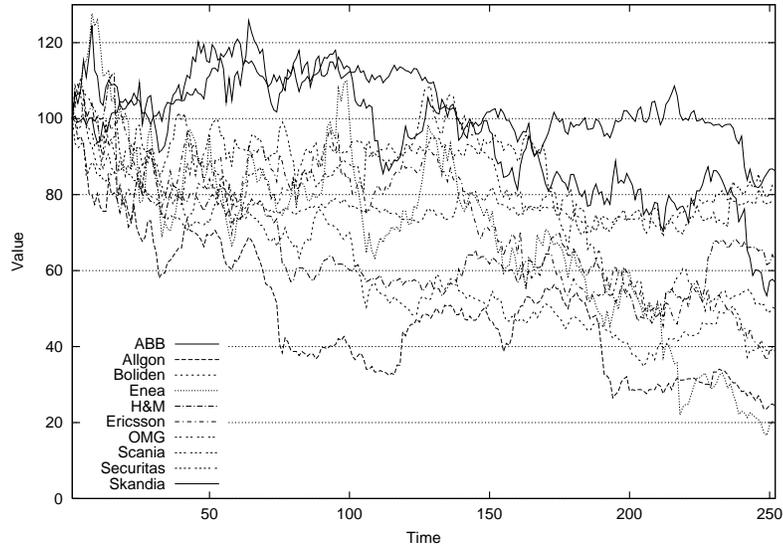,width=0.9\linewidth}
\caption{The development of the values of the stocks used in the experiment.}
\label{stockfig}
\end{figure}

We considered a portfolio of ten stocks with receding prices, assumed to be unaffected by agent trading. The data used is real daily data from the Swedish stock market, from the one-year period starting March 1, 2000. The stocks are listed in Table~\ref{stocktable}, and in Figure~\ref{stockfig} their development is shown. Values have been normalized to 100 for the start of the period. The strategies initially held \$10000 value of each stock. One trade was done per day, in which the strategy decided what to sell and what to reinvest in. Three different levels of {\em hint probabilities} were used: 1\%, 5\%, and 10\% chance of receiving a hint. A 1\% level means that the strategy will on average receive a hint for one of the ten stocks every tenth day of trading. When choosing randomly what to buy and what to sell, 10 integers were randomized and taken modulo 10 in order to get (at most 10) stocks that were then traded. For each of the stocks sold, a percentage of the possession $p \in [0.2,0.8]$ was sold. The values of all sales were then reinvested according to their relative part in a similar selection process. If the strategy did not get at least one stock to buy \emph{and} one to sell, it held its possessions until the next day. Each strategy was evaluated towards the same set of stocks and the same set of hints (if used). In order to even out differences due to the randomness of the trading, the simulations were repeated 1000 times. Alignment and docking experiments are encouraged, and specifics are available upon request.

\begin{table}
\begin{center}
\begin{tabular}{|lll|}
\hline
Stock & Business area &Value\\
\hline
\hline
ABB & Industrial &83.33\\
Allgon & Telecom &24.55\\
Boliden & Mining &37.19\\
Enea Data& IT &20.09\\
Hennes\&Mauritz & Clothes &60.40\\
Ericsson & Telecom &36.36\\
OM & Financial &48.67\\
Scania& Industrial &77.80\\
Securitas& Security &80.35\\
Skandia& Insurance &53.22\\
\hline
\end{tabular}
\end{center}
\caption{The ten stocks used in the experiment, and their normalized values on March 1, 2001.}
\label{stocktable}
\end{table}

\section{Experiment Results}
\label{simulation}

\begin{figure}
\epsfig{file=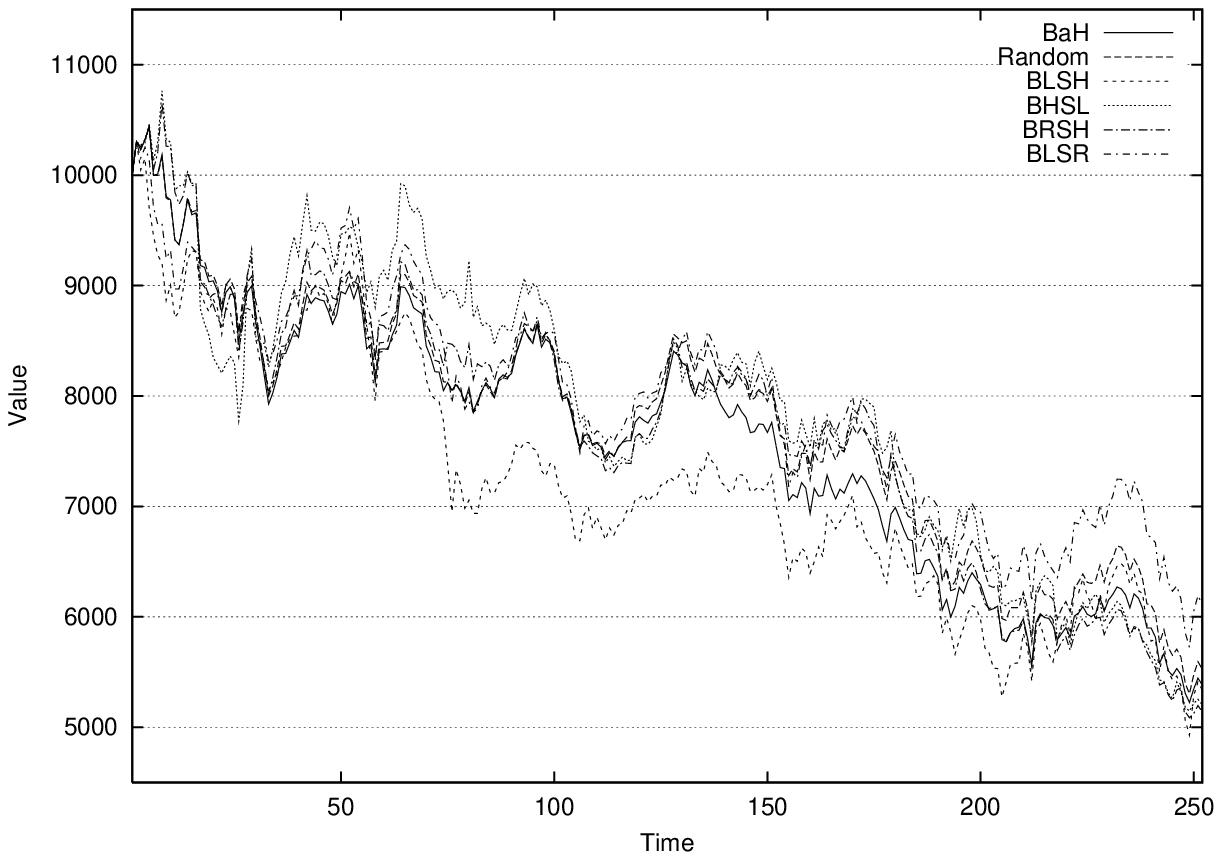,width=0.9\linewidth}
\caption{The development of the values of the experiment portfolios.}
\label{apostfig}
\end{figure}

\begin{figure}
\epsfig{file=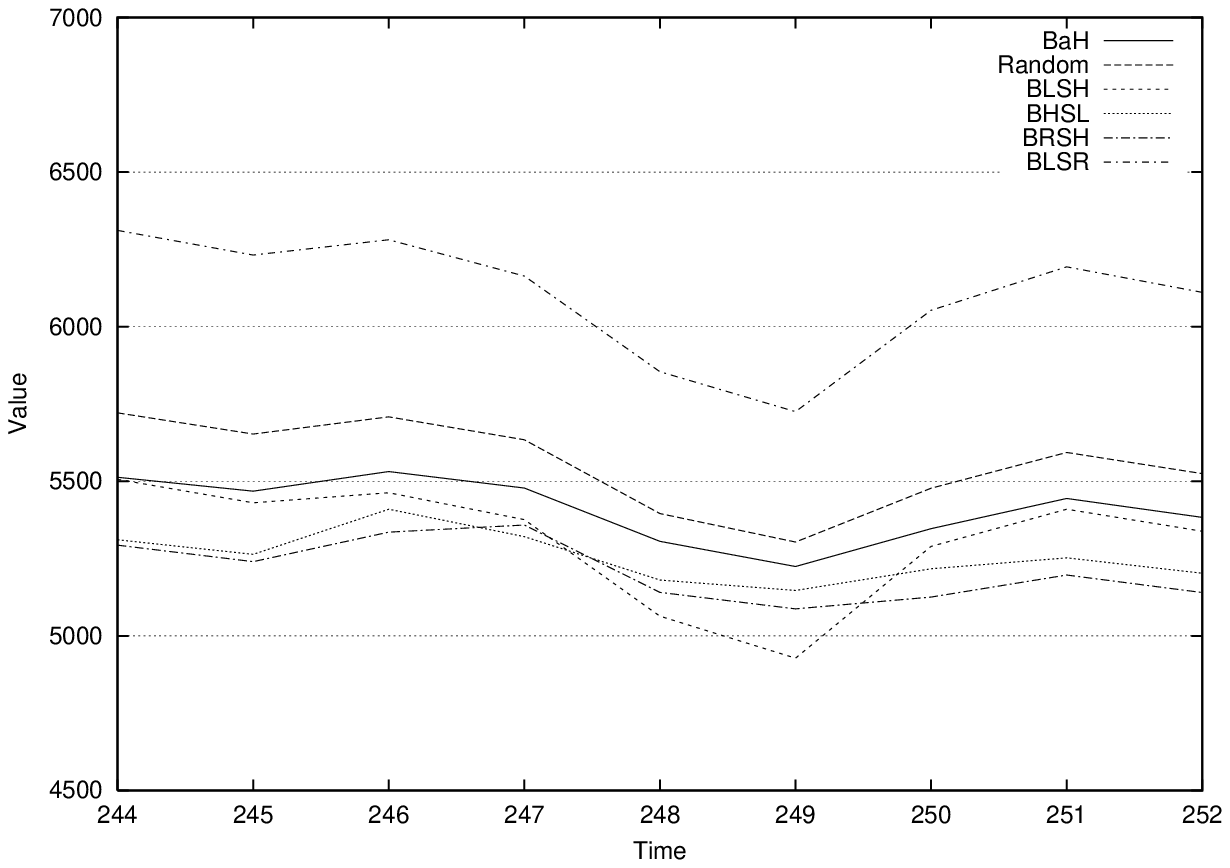,width=0.9\linewidth}
\caption{Magnification of the last days of trading.}
\label{ablowupfig}
\vspace{-10mm} 
\end{figure}

As can be seen in Figure~\ref{apostfig}, most of the strategies over the 252 trading days followed the major trends of the market and none of them managed to maintain the initial portfolio value. There was considerable movement, as shown in the blowup of the last days of trading in Figure~\ref{ablowupfig}, but also significant differences between outcomes (Table~\ref{resulttable}). Buy-low-sell-random was the only strategy that outperformed Random. Strategies also differed with respect to volatility. For instance, BLSH was inferior to all strategies for most of the year. However, around day 100 through day 120, it outperformed all other strategies. As expected, BHSL amplified the receding trend. 

\begin{figure}
\epsfig{file=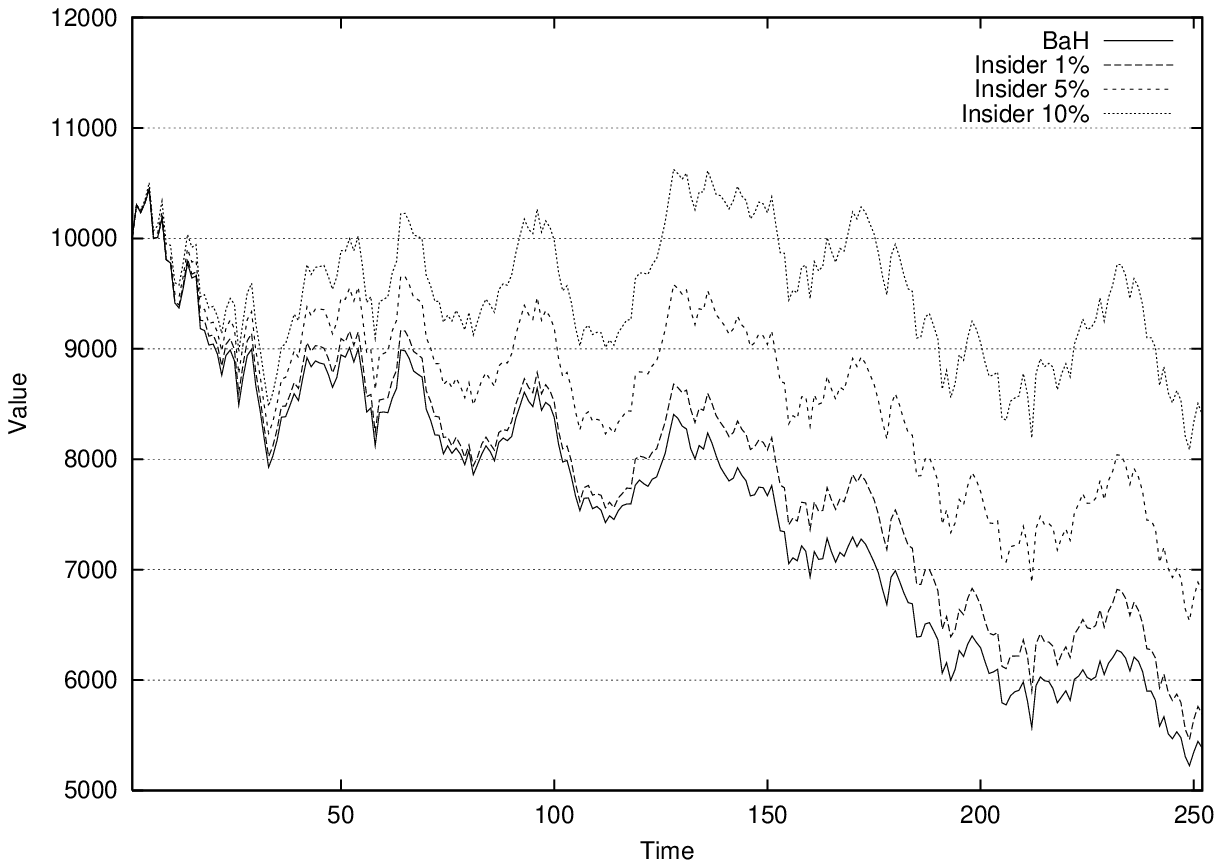,width=0.9\linewidth}
\caption{The development of the values with three different levels of hint probabilities.}
\label{aprefig}
\end{figure}

In spite of its poor performance, there are still many reasons for policy makers and speculators to use buy-and-hold even on supposedly receding markets. One reason is to declare and uphold a clear company investment policy, another is that frequent re-investments could be undesirable (e.g., due to transaction fees). Nevertheless, we feel that BLSR produced good enough results to merit further study. For now, we will be content with comparing it to various levels of hint probabilities, however. From those results, shown in Figure~\ref{aprefig}, we see that BLSR is comparable to the insider strategy with a hint probability of approximately 4\%.

\begin{table}
\begin{center}
\begin{tabular}{|p{0.18\linewidth}|p{0.12\linewidth}|}
\hline
Strategy & Value\\
\hline
\hline
BLSR & 6110.88\\
\hline
Random & 5524.60\\
\hline
BaH & 5383.40\\
\hline
BLSH & 5338.15\\
\hline
BHSL & 5202.71\\
\hline
BRSH & 5140.29\\
\hline
\end{tabular}
\end{center}
\caption{Strategy results without hint probabilities (strategies are explained in Table \ref{strategytable}).}
\label{resulttable}
\end{table}

\section{Conclusions and Directions for Future Research}
We have shown that the use of certain Parrondo-based strategies may improve the performance of artificial traders.
Our model is simplistic, in the following respects. The messages sent must be allowed to have richer content, and may be indicators or signals, rather than simple instructions. Instead of interpreting received messages as normative advice, trust could somehow be represented. For instance, a probability distribution may be associated with messages, and trust assignments can then be represented as second-order probabilities. Market norms should be modeled and adhered to by the traders\cite{bo99}. Message content can then depend on market dynamics. Artificial traders have two ways of communicating such dynamics. Firstly, they may observe and recognize other traders and try to model them with the intent of communication and possibly co-operation\cite{bo01}. Secondly, they may monitor prices, as in the Trading Agent Competition\cite{bo01a} (see \url{tac.eecs.umich.edu/}) or artificial stock market approaches\cite{le00}. Naturally, each trader itself also observes the market dynamics. We have placed no reasoning facilities in the trader at this stage, and so the trader cannot act on sense data. Another simplification is that our models should incorporate transient phenomena, including not only crashes and bubbles, but also transient diversity, i.e. we must find the homogeneity and heterogeneity drivers in our MAS\cite{ly99}. A related point in need of further investigation is learning in artificial traders\cite{le97}. 

For purposes of prediction, our results are almost useless, since we cannot in general design in advance a portfolio of stocks, the prices of which are all receding. In rare circumstances, such as during the period of almost universally receding prices of IT stocks in the autumn of 2000, ex ante portfolios could relatively easily be assembled, and then Parrondo variations would indeed be an interesting alternative to buy-and-hold. For our experiment, the real data was chosen ex post from a large sample space with the criterion that each stock should have a saw-tooth receding price curve.

While the above shortcomings together render our results useless for practical purposes, they should be seen as directions for future research. We intend to pursue the important question of strategy programming for artificial traders, as we feel that such programming will be of increasing importance in the future. By replacing our unrealistic assumptions one by one, we hope to achieve our ultimate goal of reasonably efficient strategies on real-time markets with non-linear dynamics.

\section*{Acknowledgements}
Magnus Boman was in part financed by a NUTEK (VINNOVA) grant within the {\em PROMODIS (Programming modular and mobile distributed systems)} programme. Stefan J. Johansson was financed by the KK-foundation. David Lyb\"ack was supported by a research assignment in the OM corporation. The authors wish to thank Fredrik Liljeros, as well as their respective colleagues, for comments on drafts.

\end{document}